\documentclass[prb,twocolumn,superscriptaddress,showpacs,amscd,amsmath,amssymb,verbatim]{revtex4}

\usepackage{graphicx,color}
\definecolor{red}{rgb}{1,0,0}

\usepackage{epstopdf}
\usepackage{textcomp}
\usepackage[OT1,T1]{fontenc}

\begin{document}
\newcommand{\ves}{vesignieite}
\newcommand{\herb}{herbertsmithite}

\title{Dzyaloshinsky-Moriya interaction in vesignieite:\\
A route to freezing in a quantum kagome antiferromagnet}
\author{A. Zorko}
\affiliation{Jo\v{z}ef Stefan Institute, Jamova c.~39, 1000 Ljubljana, Slovenia}
\affiliation{EN--FIST Centre of Excellence, Dunajska c.~156, SI-1000 Ljubljana, Slovenia}
\email{andrej.zorko@ijs.si}
\author{F. Bert}
\affiliation{Laboratoire de Physique des Solides, Universit\'e Paris-Sud 11, UMR CNRS 8502, 91405
Orsay, France}
\author{A. Ozarowski}
\affiliation{National High Magnetic Field Laboratory, Florida State University, Tallahassee, Florida 32310, USA}
\author{J. van Tol}
\affiliation{National High Magnetic Field Laboratory, Florida State University, Tallahassee, Florida 32310, USA}
\author{D. Boldrin}
\affiliation{Department of Chemistry, University College London, 20 Gordon Street, London WC1H 0AJ, UK}
\author{A. S. Wills}
\affiliation{Department of Chemistry, University College London, 20 Gordon Street, London WC1H 0AJ, UK}
\author{P. Mendels}
\affiliation{Laboratoire de Physique des Solides, Universit\'e Paris-Sud 11, UMR CNRS 8502, 91405 Orsay, France}
\affiliation{Institut Universitaire de France, 103 Boulevard Saint-Michel, F-75005 Paris, France}
\date{\today}
\begin{abstract}
We report an electron spin resonance investigation of the geometrically frustrated spin-1/2 kagome antiferromagnet vesignieite, BaCu$_3$V$_2$O$_8$(OH)$_2$. 
Analysis of the line widths and line shifts indicates the dominance of in-plane Dzyaloshinsky-Moriya anisotropy that is proposed to suppress strongly quantum spin fluctuations and thus to promote long-range ordering rather than a spin-liquid state. We also evidence an enhanced spin-phonon contribution that might originate from a lattice instability and discuss the origin of a low-temperature mismatch between intrinsic and bulk susceptibility in terms of local inhomogeneity.
\end{abstract}
\maketitle

\section{Introduction}
The two-dimensional spin-1/2 quantum kagome antiferromagnet (QKA) has been early recognized as an ideal candidate for stabilizing a spin-liquid state.\cite{Balents, LMM} The possibility of generating fractionalized excitations such as spinons and the nature itself of its ground state (GS) have been hotly debated over the last 20 years with proposals of many competing states -- gapped\cite{Yan,Depenbrock} and gapless\cite{Ran,Iqbal} spin liquids, as well as valence-bond solids.\cite{Singh,Evenbly} While the most recent calculations clearly point to a gapped spin-liquid GS,\cite{Yan,Depenbrock} likely a resonating valence bond (RVB) state, the few experimental realizations have been found gapless,\cite{Mendels,Clark} thus apparently contradicting this scenario. 

It is commonly advocated that the issue of opposing experimental findings and theoretical predictions results from weak perturbing interactions in the context of a GS manifold of the isotropic Heisenberg exchange. The most deeply studied case is that of the out-of-plane Dzyaloshinsky-Moriya (DM) magnetic anisotropy\cite{DM} $D_z$ that is present in any real QKA as the bonds lack inversion symmetry and is theoretically predicted to create a quantum critical point (QCP) at $D_z^c/J\simeq0.1$.\cite{Cepas} This separates a moment-free phase ($D_z<D_z^c$) from a N\'eel-ordered phase ($D_z>D_z^c$).\cite{Cepas,Messio,Huh} The mineral \herb, $\gamma$-ZnCu$_3$(OH)$_6$Cl$_2$, believed so far to be the best realization of the QKA,\cite{Shores} appears to be a spin liquid \cite{Mendels,deVries2} sustaining spinon excitations,\cite{HahnNature} in line with its DM anisotropy $D_z/J=0.06(2)$.\cite{Zorko,Rousochatzakis,Samir} Its location in the region close to the QCP is likely responsible for observed field- \cite{Jeong} and pressure-induced \cite{Kozlenko} freezing. This theoretical scenario awaits further validation, potentially by finding new compounds lying in the N\'eel-ordered region of the phase diagram.
\begin{figure}[b]
\includegraphics[trim = 17mm 13mm 58mm 23mm, clip, width=1\linewidth]{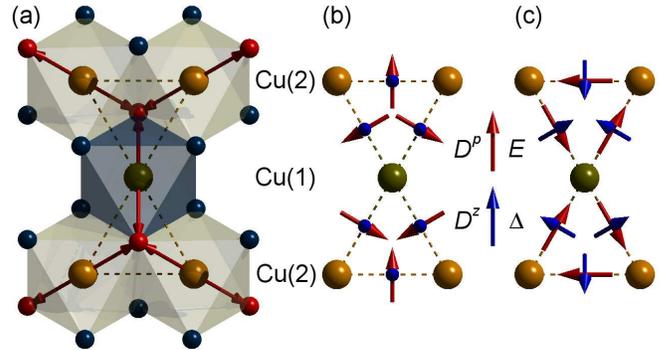}
\caption{The two inequivalent copper sites Cu(1) and Cu(2) on the kagome lattice in  \ves~($ab$ crystallographic plane). (a) The double-headed arrows connect apical O(1) sites in each CuO$_6$ octahedron and define the principal axis of the $g$-tensor on each Cu$^{2+}$ site. (b) The Dzyaloshinsky-Moriya (DM) pattern of out-of-plane $D_z$ (uniform) and in-plane $D_p$ components. (c) Two principal directions, $\Delta$ and $E$, of the local symmetric anisotropic-exchange (AE) tensor; $\Delta$ is canted by $\theta_0=45^\circ$ out of the kagome plane, while $E$ lies in the plane.}
\label{Fig1}
\end{figure}

In this context, the mineral vesignieite, BaCu$_3$V$_2$O$_8$(OH)$_2$, which has been recently highlighted as a new realization of the QKA,\cite{Okamoto} is an appealing case. It crystallizes in the monoclinic space group\cite{Lafontaine} $C2/m$ and the minute 0.07\% bond-length difference due to two inequivalent Cu$^{2+}$ sites (Fig.~\ref{Fig1}) makes the triangles very close to being equilateral.\cite{Colman} Indeed, there have been suggestions that the actual structure has equilateral symmetry,\cite{Yoshida} though this may not yet be conclusive.\cite{Boldrin} The magnetism of vesignieite is dominated by the nearest-neighbor antiferromagnetic interaction\cite{Okamoto} $J=53$~K that leads to a maximum in local susceptibility $\chi_i$ at the temperature $T\simeq 0.5J$,\cite{Quilliam} detected by nuclear magnetic resonance (NMR).
In marked contrast to \herb, vesignieite shows a magnetic transition\cite{Colman,Yoshida,Boldrin,Quilliam,MYoshida} to a $q=0$ N\'eel state at $T_N=9$~K. During  this spin freezing transition, an additional out-of-plane spin component creates a ZFC/FC bifurcation.
Based on the width of electron-spin-resonance (ESR) spectra\cite{Zhang} vesignieite has been suggested to possess large DM anisotropy $D_z$ and thus to be in the ordered region of the phase diagram.\cite{Colman,Quilliam,Yoshida, MYoshida} Since vesignieite appears to be the first clear case of a long-range ordered QKA and no proper attempt to identify and quantify its magnetic anisotropy has been reported, a detailed study is essential.

In this paper, we clarify the driving force of magnetic ordering in vesignieite by determining its dominant magnetic anisotropy. Employing the local-probe ESR technique we show that the in-plane component of the dominant DM anisotropy, $D_p$, exceeds $D_z$, in contrast to herbertsmithite. We propose that such a DM vector crucially suppresses quantum fluctuations and thus critically affects the GS of this material. Additionally, we assess the importance of a symmetric anisotropic exchange (AE) that has recently been suggested as an important spin-Hamiltonian component of herbertsmithite.\cite{Han, Ofer}

\section{Experimental details}
Our ESR experiments were conducted at 328.8~GHz on a custom-made spectrometer working in transmission mode at the NHMFL, Tallahassee, USA, 
allowing single field-sweep detection of spectra with negligible background.
The sample was hydrothermally annealed powder similar to that used in previous studies.\cite{Colman,Quilliam}

\section{Theoretical background}
ESR has proven extremely efficient for determining magnetic anisotropy, either by detecting collective excitations,\cite{ZorkoFe,Zvyagin} or through the modeling of shifts\cite{Povarov,Furuya} and line widths\cite{Zorko,ZorkoSCBO} of a paramagnetic resonance. Both, the shifts and widths are non-zero only when the anisotropy is finite.\cite{AB} They allow distinction of different forms of the anisotropy and its direct quantification.

In an ESR experiment a magnetic system is exposed to the applied static magnetic field $B_0$ and an electromagnetic wave with polarization of its magnetic field perpendicular to $B_0$ (conventional Faraday configuration). Within the linear response theory the absorption spectrum $I(\omega)$ is proportional to the imaginary part of the dynamical susceptibility,\cite{KT}
\begin{equation}
I(\omega)\propto\chi"({\rm \textbf{q}}\rightarrow0,\omega) \propto \int_{-\infty}^{\infty}{\rm d}t \left\langle S^+(t)S^-(0)\right\rangle \exp^{i \omega t}/T,
\label{eqs1}
\end{equation}
and thus effectively measures spin correlations in the direction perpendicular to the applied field, $\left\langle S^+(t)S^-(0)\right\rangle$, where $\langle\;\rangle$ denotes canonical averaging and $S^\alpha=\sum_iS_i^\alpha$ is the $\alpha$-component of the total spin operator.

Calculating the time-dependent spin operator in the Heisenberg representation $S^+(t)={\rm e}^{\frac{i}{\hbar}\mathcal{H}t}S^+{\rm e}^{-\frac{i}{\hbar}\mathcal{H}t}$ for a general spin Hamiltonian $\mathcal{H}$ is a nontrivial problem. Therefore, a few approximate solutions have been developed. The well-established Kubo-Tomita (KT) approach\cite{KT} relies on dividing the spin Hamiltonian into two parts, $\mathcal{H}=\mathcal{H}_0+\mathcal{H}'$, where the first, dominant part $\mathcal{H}_0$ contains only the Zeeman term and the Heisenberg isotropic exchange ($J$), while the magnetic anisotropy term $\mathcal{H}'$ is treated as a perturbation. The latter then determines the shape of the ESR spectrum (its position and width), because $\mathcal{H}_0$ possessing SU(2) symmetry conserves the total magnetization and therefore leads to a $\delta$-function resonance at the field $B_0$. For finite $\mathcal{H}'$, at high temperatures ($T\gg J$) the KT theory predicts a Lorentzian exchange-narrowed\cite{Anderson} ESR absorption line with the full-width-half-maximum line width\cite{Castner}
\begin{equation}
\Delta B = C \frac{k_B}{g \mu _B} \sqrt{\frac{M_2^3}{M_4}},
\label{eqs2}
\end{equation}
where 
\begin{align}
M_{2}&= \notag \frac{\left\langle \left[ \mathcal{H}^{\prime },S^{+}\right]
[ S^{-},\mathcal{H}^{\prime }] \right\rangle} {\left\langle S^{+}S^{-}\right\rangle},\\
M_{4}&=\frac{\left\langle \left[ \mathcal{H}-\mathcal{H}_{Z},\left[
\mathcal{H}^{\prime },S^{+}\right] \right] [ \mathcal{H}-\mathcal{H}_{Z},\left[
\mathcal{H}^{\prime },S^{-}\right] ] \right\rangle}{ \left\langle
S^{+}S^{-}\right\rangle},
\end{align}
are the second and the fourth moment of the absorption line, respectively, with $\left[ \; \right]$ denoting a commutator, $C$ is a constant of the order of unity (see below), $k_B$ stands for the Boltzman constant and $\mu_B$ for the Bohr magneton. The expression~(\ref{eqs2}) is valid if the magnetic anisotropy is small compared to $\mathcal{H}_0$ and if spin diffusion is negligible, which is generally the case in spin systems with dimensionality exceeding one.\cite{Richards} Strictly speaking, the ESR absorption line is never truly Lorentzian, as all of its moments, given by the spin Hamiltonian, are always finite while they diverge for the Lorentzian line shape. In systems with strong isotropic exchange compared to magnetic anisotropy deviations from the Lorentzian shape occur only in far wings of the resonance and an approximate line shape that is a product of the Lorentzian and a broad Gaussian $\propto{\rm e}^{-(B-B_0)^2/2B_e^2}$,\cite{Castner} with $B_e=k_B/g\mu_B\sqrt{M_4/M_2}$ being the exchange field, is applicable. This then yields $C=\sqrt{2\pi}$. 

\section{Results}
\begin{figure}[t]
\includegraphics[trim = 0mm 22mm 0mm 15mm, clip, width=1\linewidth]{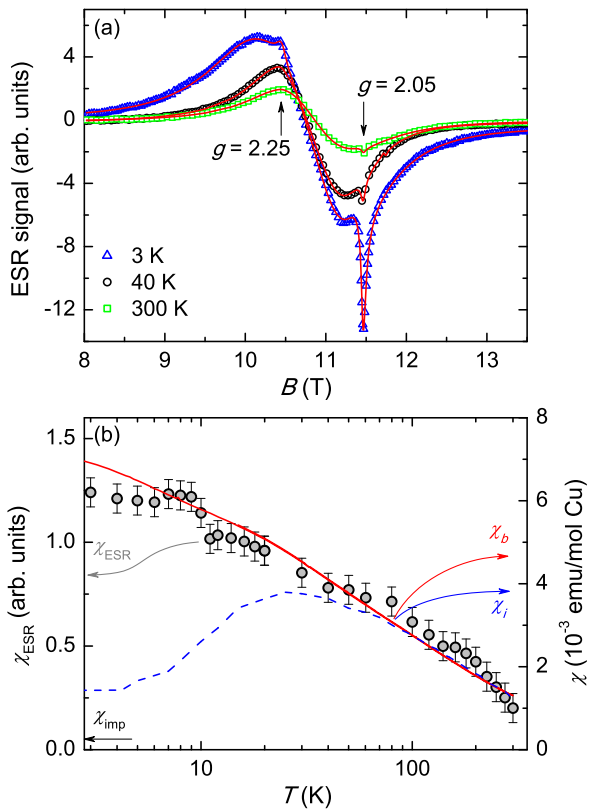}
\caption{(a) The ESR spectra (symbols) measured at 328.8~GHz and the fits (lines). Arrows point to impurity lines with corresponding $g$-factors. (b) Comparison of the ESR intensity $\chi_{\rm ESR}$ (symbols) with bulk magnetic susceptibility $\chi_b$ (solid line) measured in 5~T and the intrinsic susceptibility $\chi_i$ (dashed line) obtained from $^{51}$V NMR.\cite{Quilliam} The arrow indicates the intensity of the impurity ESR signal at 3~K.}
\label{Fig2}
\end{figure}

In Fig.~\ref{Fig2}(a) we show derivative ESR spectra typical of those recorded in the $T$-range between 3 and 300~K. The spectra have similar width as in \herb,\cite{Zorko} suggesting that substantial magnetic anisotropy is also present in \ves. In addition to this broad component, we observe a narrow component with the principal $g$-factors values of 2.05 and 2.25, typical of Cu$^{2+}$ ions.\cite{AB} We attribute this narrow component to a minor impurity phase since its intensity at 300~K amounts to only 0.3\% of the broad-component intensity and exhibits a Curie-like $T$-dependence. The impurity signal is thus much too small to explain the substantial low-$T$ increase of the bulk susceptibility $\chi_b$ [Fig.~\ref{Fig2}(b)]. The ESR intensity of the broad component, $\chi_{\rm ESR}$, convincingly follows $\chi_b$ [Fig.~\ref{Fig2}(b)], and not the nonmonotonic intrinsic $\chi_i$. This has an important implication for the hitherto unknown origin of the low-$T$ increase of $\chi_b$.\cite{Okamoto,Colman} Indeed, the observation of a single broad-component ESR line, rather than two distinct lines, reveals that the spins contributing to $\chi_b$ are necessarily exchange coupled with the intrinsic Cu$^{2+}$ spins. Bond disorder due to oxygen vacancies or some other non-stoichiometry effect then provides a credible explanation for the mismatch between $\chi_b$ and $\chi_i$. Such disorder also explains the inhomogeneous broadening of the $^{51}$V NMR lines far above $T_N$.\cite{Quilliam}

For the magnetic field applied perpendicular ($z$) and within the kagome plane ($p$) we find $g_z>g_p$. As the local anisotropy axis, set by the direction of the the shortened Cu-O(1) bond [Fig.~1(a)], makes the angle of $26.7^\circ<\pi/4$ with the kagome plane, the principal $g$-factor value $g_\|$ along the apical direction is smaller than the value $g_\bot$ in the perpendicular direction. This confirms that the Cu$^{2+}$ orbital state involves occupation of the $d_{3z^2-r^2}$ state\cite{Okamoto} rather than the more common $d_{x^2-y^2}$ orbital that would lead to $g_\|>g_\bot$.\cite{AB}

\subsection{ESR line-width analysis}
In order to determine the $T$-evolution of the ESR line width and its origin we first fitted the experimental spectra [Fig.~\ref{Fig2}(a)] to a powder-averaged line shape based on a field distribution originating from the $g$-factor anisotropy $g(\theta)=(g_z^2{\rm cos}^2\theta+g_p^2{\rm sin}^2\theta)^{1/2}$ that is convoluted with a Lorentzian function with the phenomenological 
line width $\Delta B(\theta)=(\Delta B_z^2{\rm cos}^2\theta+\Delta B_p^2{\rm sin}^2\theta)^{1/2}$. Here $\theta$ denotes the polar angle between the applied magnetic field and the normal to the plane. The presumed independence of both $g$ and $\Delta B$ on the azimuthal angle stems from the near threefold rotational symmetry of the lattice. Our approach elaborates on the previous ESR report that employed a simpler analysis yielding only an average $\Delta B$ and $g$-factor.\cite{Zhang} 

\begin{figure}[t]
\includegraphics[trim = 0mm 22mm 0mm 15mm, clip, width=1\linewidth]{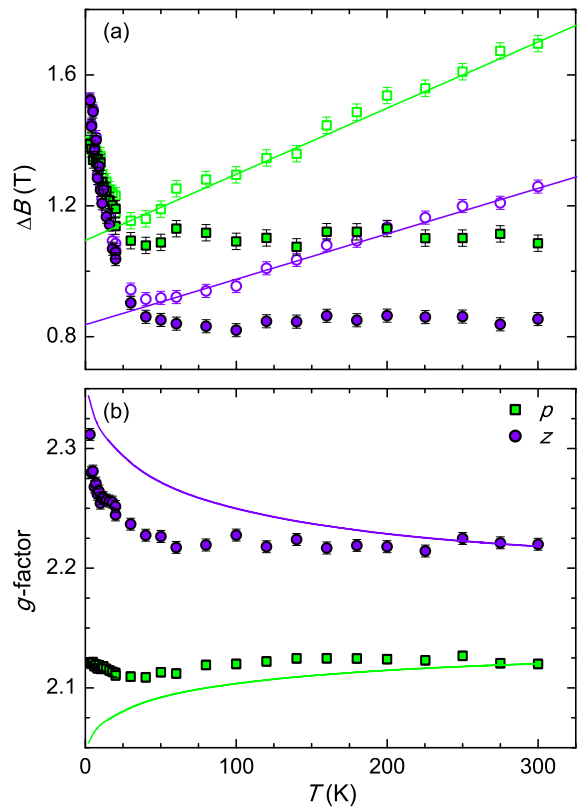}
\caption{(a) $T$-dependence of the ESR line width (open symbols) for the perpendicular ($z$) and in-plane ($p$) directions of the applied magnetic field. Linear contribution (lines) is subtracted to obtain intrinsic kagome-lattice line width (full symbols). (b) The anisotropic-exchange model leads to a  sizable temperature variation of the $g$-factor (lines) which is not detected in the experimental data (symbols).}
\label{Fig3}
\end{figure}

The ESR line width in \ves~exhibits a minimum at $T_{\rm min}=40$~K and increases linearly with $T$ at least up to 300~K, i.e., $T/J\sim 6$ [Fig.~\ref{Fig3}(a)]. This is in sharp contrast to \herb~where it was found constant for $T\gtrsim J$.\cite{Zorko} In general, dying out of spin correlations above the characteristic exchange temperature $J$ causes a vanishing contribution to the ESR line width. Similar linearly increasing ESR line width at surprisingly high temperatures ($T\gg J$) was observed in localized-spin systems  on several instances.\cite{ZorkoSCBO, Castner, Seehra, HuberJPCS, SeehraJPCM, Huber, Heinrich,  Deisenhofer} Such behavior can arise either from the phonon modulation of the anisotropic exchange\cite{Seehra} or the crystalline field, the latter for $S>1/2$.\cite{HuberJPCS} 
We therefore attribute the observed behavior in \ves~to an additional line-broadening mechanism that arises from a spin-phonon coupling and is due to a direct phonon process yielding the linearly increasing ESR relaxation due to phonon modulation of magnetic anisotropy.\cite{Seehra} We propose that the linear increase might be related to a structural instability of \ves~associated with the energetic proximity of the monoclinic crystal structure to the higher-symmetry rhombohedral ($R\bar{3}m$) structure, the latter with a perfectly undistorted kagome lattice.\cite{Yoshida}

In cases of the phonon-induced ESR broadening the line width is regularly written as a sum of a temperature independent and a linearly increasing contribution. Such division is justified for relaxation mechanisms that contribute independently to the relaxation of the spin correlation function $\left\langle S^+(t)S^-(0)\right\rangle$, whose decay is close to being exponential (for close-to Lorentzian line shapes), as regularly encountered in concentrated magnetic insulators. The usual subtraction of the linearly increasing line-width contribution in \ves~gives the width intrinsic to the kagome spin system. Its increase with decreasing $T$ below $T_{\rm min}$ is similar to that observed in \herb~and can be attributed to the building-up of spin correlations, which are also responsible for the maximum in $\chi_i$ at 25~K.\cite{Quilliam} In order to determine the magnetic anisotropy, we therefore make use of the 40~K spectrum, which corresponds well to the paramagnetic limit, as both the spin-phonon and the spin-correlation induced broadenings are small. Moreover, since $\chi_{\rm ESR}$ is not much different from $\chi_i$ at 40~K we do not expect any notable effect of the bond disorder on the 40~K ESR spectrum.\cite{Huber2}

In this paramagnetic limit, we model the ESR line-width anisotropy $\Delta B(\theta)$ by employing the Kubo-Tomita moment approach [Eq.~(\ref{eqs2})]. Both, the DM magnetic anisotropy [Fig.\ref{Fig1}(b)] 
\begin{equation}
\mathcal{H'_{\rm DM}}=\sum_{(ij)}{\bf D}_{ij}\cdot {{\bf S}_{i} \times {\bf S}_{j}}
\label{DM} 
\end{equation}
and the traceless symmetric anisotropic exchange [Fig.\ref{Fig1}(c)], written in a local basis as 
\begin{eqnarray}
\mathcal{H'_{\rm AE}}&=\sum_{(ij)} \Big[ \frac{2\Delta}{3} S_i^\xi S_j^\xi
+\left(-\frac{\Delta}{3}+\frac{E}{2} \right)S_i^\eta S_j^\eta \nonumber \\
&+\left(-\frac{\Delta}{3}-\frac{E}{2} \right)S_i^\nu S_j^\nu \Big],
\label{AE}
\end{eqnarray}
yield the line width of the general form
\begin{equation}
\Delta B(\theta) = \sqrt{2 \pi} \frac{k_b}{2 g(\theta) \mu_B J}
\sqrt{ \frac{(a+b\;{\rm cos}^2\theta)^3}{c+d\;{\rm cos}^2\theta}},
\label{eq1}
\end{equation}
where $g(\theta)$ denotes the $g$-factor averaged over the basic hexagon of the kagome lattice, and the constants $a$, $b$, $c$, $d$ are related to the anisotropy constants of each model [see  Eqs.~(\ref{DMwidth}),~(\ref{AEwidth}) in the Appendix~\ref{appA}).
Although this angular dependence is more complicated than the phenomenological one employed above, their differences are minimal (see Appendix~\ref{appB}), assuring that the $T$-dependences of all ESR parameters in Fig.~\ref{Fig3} are meaningful.
\begin{figure}[t]
\includegraphics[trim = 00mm 9mm 0mm 6mm, clip, width=1\linewidth]{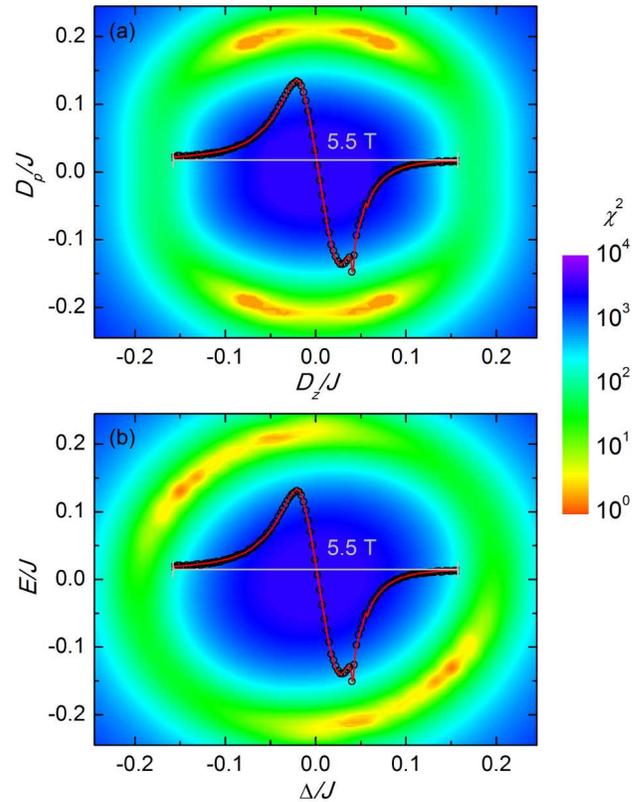}
\caption{Reduced $\chi^2$ of fitting the 40~K ESR spectrum with (a) the DM and (b) the AE model. The former yields optimal parameters $|D_p|/J =0.19(2)$, $|D_z|/J =0.07(3)$ and the latter two inequivalent solutions, $\Delta/J =\pm 0.15(2)$, $E/J=\mp 0.13(2)$ and $\Delta/J=\pm 0.04(2)$, $E/J=\mp 0.21(1)$. Center of each panel: Comparison of the best fit (line) and experimental data.}
\label{Fig4}
\end{figure}

Fitting the experimental spectrum to the powder-averaged line shape with the line width given by Eq.~(\ref{eq1}) provides fits of equal quality for both models. These are displayed in Fig.~\ref{Fig4} together with $\chi^2$ maps spanned over the parameter space. For the DM model, we find the solution $|D_p|/J =0.19(2)$, $|D_z|/J =0.07(3)$, while the AE model yields two inequivalent solutions, $\Delta/J =\pm 0.15(2)$, $E/J=\mp 0.13(2)$ and $\Delta/J=\pm 0.04(2)$, $E/J=\mp 0.21(1)$. The relative sizes of the DM and AE anisotropy are very similar with respect to the dominant exchange $J$. Since the DM interaction results from a first order correction of $J$ in the spin-orbit coupling, which is for the Cu$^{2+}$ ions a $\sim$10\% perturbation on $J$,\cite{AB} while the AE interaction is a second-order correction, the DM interaction is generally considered dominant. However, caution is necessary as $D_p$, the largest anisotropy in the DM model in \ves, is reducible on the kagome lattice.\cite{Cepas} This is because it possesses a hidden symmetry\cite{Shekhtman} and can be transformed into an effective term of the order $D_p^2/J$ by applying a nonuniform spin rotation.\cite{Choukroun} Therefore, we provide a second criterion that is based on the ESR line shift, which allows distinction between the two anisotropy models in \ves.

\subsection{ESR line-shift analysis}
When the anisotropy is smaller than the isotropic exchange, as is the case here, Nagata's theory\cite{Nagata} of the ESR line shift can be applied. Accordingly, the shift of the $g$-factor from its infinite-$T$ value $g^\infty$ is given by the first moment,\cite{Nagata,Maeda}
$g-g^\infty = {\left\langle \left[ S^-,\left[S^+,\mathcal{H}' \right] \right] \right\rangle}/{2\mu_B B_0\left\langle S^z\right\rangle}$.
It is important to stress that in this first-order calculation (in $\mathcal{H}'$) the DM interaction leads to zero shift,\cite{Maeda} while the shift due to the AE interaction scales with the susceptibility $\chi$ in the paramagnetic regime,\cite{Nagata,Nagata2}
\begin{equation}
g_Z-g_Z^\infty = \frac{\chi}{2N_A g \mu_0 \mu_B^2}\sum_{j\neq i}{\left(2 \Delta_{ij}^{ZZ}-\Delta_{ij}^{XX}-\Delta_{ij}^{ZZ}\right)}.
\label{shift2}
\end{equation}
Here $\Delta^{ZZ}$ denotes the component of the AE tensor along the applied field and $\Delta^{XX},\,\Delta^{YY}$ in two perpendicular directions, $N_A$ is the Avogadro number and $\mu_0$ the vacuum permeability. Quantifying this expression in \ves~with the above-determined AE parameters, we arrive at the scalings $(g_z-g_z^\infty)/\chi = \pm 22.3$~(emu/mol~Cu)$^{-1}$ and $(g_p-g_p^\infty)/\chi = \mp 11.7$~(emu/mol~Cu)$^{-1}$ for the two relevant directions. These values significantly overestimate the measured ESR shifts that are found constant for $T\gtrsim 50$~K within much smaller error bars [Fig.~\ref{Fig3}(b)]. The $g^z$ data limit the AE anisotropy to at least 4-times smaller values. Therefore, the ESR line width, which scales with the square of the anisotropy, is entirely determined by the dominant DM interaction.

The experimental ESR shift becomes sizable below 25~K [Fig.~\ref{Fig3}(b)], which was ascribed  previously to short-range ordering effects.\cite{Zhang} However, due to the traceless nature of the AE interaction it is generally expected that even in the short-range correlated regime the ESR shift will exhibit both positive and negative shifts for different directions of the applied field,\cite{Nagata,Nagata2} just as occurs at higher $T$. This is not the case, therefore, we attribute the solely positive low-$T$ $g$-shifts to local inhomogeneity present in \ves. Namely, in inhomogeneous systems (e.g., impure systems and systems with several inequivalent sites) spatially varying local fields exclusively lead to positive $g$-shifts\cite{Malozemoff,Nagata3} that stem from the $q\neq0$ Fourier components of the inhomogeneous field.\cite{Nagata3} This corroborates our  proposition, based on the scaling of $\chi_{\rm ESR}$ with $\chi_b$, that there is significant bond disorder in \ves.

\section{Discussion}
Although, in both \herb~and in \ves~the ESR spectra can be accounted for by the DM magnetic anisotropy, we point out an essential difference.
While the out-of-plane DM component $D_z$ is dominant in \herb,\cite{Zorko} it is the in-plane component $D_p$ that dominates in \ves. We additionally note that the Kubo-Tomita approach generally underestimates the strength of the reducible $D_p$ component with respect to the irreducible $D_z$ component.

Then, it is important to inspect the effect of both DM components on the GS of the QKA. Although in the classical limit both components immediately lead to magnetic ordering, their effect is rather different.\cite{Elhajal} The GS due solely to the $D_z$ component is an in-plane 120$^\circ$ spin structure invariant under a global rotation around the $z$ axis, while a finite $D_p$ component prefers a state with a finite uniform out-of-plane spin component, thus removing this state from the GS manifold of the Heisenberg Hamiltonian by eliminating the rotation symmetry. For $D_z>0$, the tilt $\phi$ of each spin from the kagome plane caused by $D_p$ is given by $\tan(2\phi)=2D_p/(\sqrt{3}J+D_z)$,\cite{Elhajal} which yields $\phi=6^\circ$ in \ves. Interestingly, the weak ferromagnetic spin component estimated by NMR\cite{MYoshida} amounts to 0.05-0.12~$\mu_B$, which together with the dominant in-plane component being larger than 0.6~$\mu_B$ limits the tilting angle to $3^\circ<\phi<9^\circ$. In the quantum picture\cite{Cepas} N\'eel ordering is induced by $D_z$ only for $D_z>D_z^c\simeq 0.1$. Including a finite $D_p$ in this case leads to a weak ferromagnetic moment in the $z$ direction that is still linear in $D_p/J$ as in the classical case,\cite{Cepas} while the position of the QCP would be affected by the $D_p^2/J$ term as well as linear terms in the AE anisotropy. In \ves, the condition $D_p>D_z$, could profoundly affect the QCP because $D_p$ disfavors spin structures from the GS manifold of the isotropic $J$ and should therefore be much more efficient in suppressing quantum fluctuations than $D_z$. This could explain why magnetic ordering in \ves~occurs at surprisingly high temperature $T_N/J=0.17$, despite possessing very similar $D_z/J$ as \herb. Comprehensive theoretical investigations of the general DM-perturbed phase diagram of the QKA are thus highly desired.

\section{Conclusions}
Employing the ESR line-width and line-shift analyses we have shown that the in-plane DM interaction is prevailing in the novel QKA \ves~and is most likely responsible for its magnetic ordering below $T_N=9$~K. We have detected intrinsic inhomogeneity of the kagome planes, which we attribute to bond disorder, as well as sizable spin-phonon contribution that might be related to a lattice instability. Last, we note that a preliminary analysis of the ESR line\cite{Zorko} of \herb~with the AE model yields anisotropy constants $|\Delta|/J =0.072$ and $|E|/J= 0.074$ that give an "effective" AE anisotropy of $|\Delta_{\rm av}|/J= 0.06$ if averaged over the triangle. This value is notably smaller than the recent estimate $\Delta_{\rm av}/J\simeq -0.1$,\cite{Han} which would lead to much broader ESR lines since their width scales with the square of the anisotropy.
Increasing the sensitivity by performing single-crystal ESR and applying the above-presented analysis
is likely the most reliable approach for resolving the standing issue of the dominant anisotropy in \herb, which could turn to be the crucial milestone in understanding its spin-liquid properties.

\acknowledgments
We thank O.~C\'epas and S.~El~Shawish for valuable discussions. AZ acknowledges the financial support of the Slovenian Research Agency (projects J1-2118, BI-US/09-12-040 and Bi-FR/11-12-PROTEUS-008). The NHMFL is supported by NSF Cooperative Agreement No.~DMR-1157490, and by the State of Florida.

\appendix

\section{ESR line-width anisotropy on the kagome lattice}\label{appA}
Within the KT theory,\cite{KT} employing the moment approach in the paramagnetic limit ($T\gg J$) when spin correlations on neighboring sites vanish, both, the Dzyaloshinsky-Moriya magnetic anisotropy [Eq.~(\ref{DM}); see Fig.~\ref{Fig1}(b)]  and the traceless symmetric anisotropic exchange interaction [Eq.~(\ref{AE}); see Fig.~\ref{Fig1}(c)], yield the ESR line width of the same general form [Eq.~(\ref{eq1})]. For the DM model, the $a$, $b$, $c$, $d$ constants are given by the two DM components $D_z$ and $D_p$ as\cite{Zorko}
\begin{align}
a &= \notag 2D_z^2+3D_p^2,\\
b &= \notag 2D_z^2-D_p^2,\\
c &= \notag 16D_z^2+78D_p^2,\\
d &= 16D_z^2-26D_p^2,
\label{DMwidth}
\end{align}
while for the AE model we derive
\begin{align}
a &= \notag 12\Delta^2-4E\Delta+7E^2+\left(4E\Delta-4\Delta^2+3E^2 \right){\rm cos}^2\theta_0,\\
b &= \notag 12E\Delta-4\Delta^2+3E^2+\left(12\Delta^2-12E\Delta-9E^2 \right){\rm cos}^2\theta_0,\\
c &= 48[ \notag 68\Delta^2-36E\Delta+33E^2+\left(72E\Delta-36\Delta^2+36E^2 \right)\notag \\
&\times {\rm cos}^2\theta_0 -\left(36\Delta^2+36E\Delta+9E^2 \right)  {\rm cos}^4\theta_0], \notag\\
d &= 48[\notag 60E\Delta-28\Delta^2+9E^2+\left(96\Delta^2-24E\Delta-36E^2 \right)\notag \\
&\times {\rm cos}^2\theta_0 -\left(36\Delta^2+36E\Delta+9E^2 \right)  {\rm cos}^4\theta_0].
\label{AEwidth}
\end{align}
Here $\theta_{\rm 0}=45^\circ$ corresponds to the angle between the $\xi$ axis of a local coordinate system and the normal to the kagome plane while the $\eta$ axis is parallel to the Cu-Cu bond. The direction of the $\xi$ and $\eta$ principal axes of the AE anisotropy tensor are determined by the sum of the $g$-factor tensors of the two sites constituting a particular bond.\cite{Bencini} The latter tensors are very close to being uniaxial, with the anisotropy axis pointing towards the apical O(1) site [see Fig.~\ref{Fig1}(a)], which leads to the above-mentioned direction of the $\xi$ and $\eta$ axes.

\section{Comparison of ESR line-width-anisotropy models}\label{appB}
\begin{figure}[h]
\includegraphics[trim = 0mm 58mm 0mm 51mm, clip, width=1\linewidth]{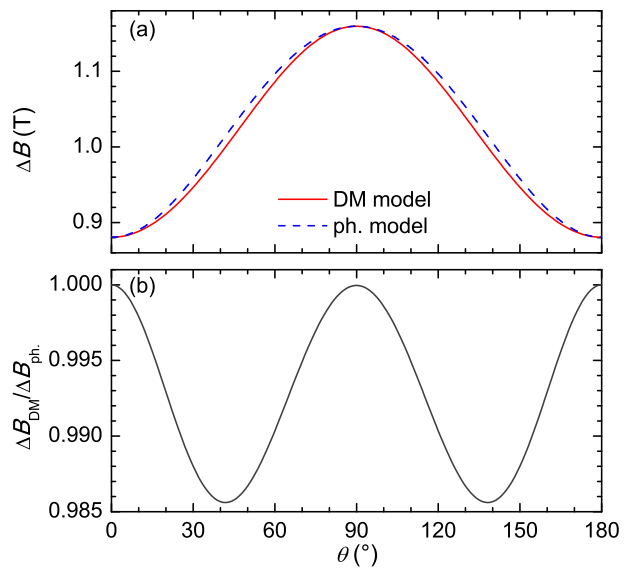}
\caption{(a) Angular dependence of the ESR line width for the DM model with the optimal parameters $|D_p|/J =0.19(2)$, $|D_z|/J =0.07(3)$ (solid line) and the phenomenological model $\Delta B(\theta)=(\Delta B_z^2{\rm cos}^2\theta+\Delta B_p^2{\rm sin}^2\theta)^{1/2}$ (dashed line). (b) The ratio of the line widths from the two models.}
\label{Fig5}
\end{figure}

The angular dependence of the line width arising from the spin Hamiltonian [Eq.~(\ref{eq1})] is more complicated than the usually presumed lowest order (in ${\rm cos}\:\theta$) phenomenological dependence $\Delta B(\theta)=(\Delta B_z^2{\rm cos}^2\theta+\Delta B_p^2{\rm sin}^2\theta)^{1/2}$. However, we find that for the optimal parameters of the DM model, or equivalently the AE model, the discrepancy of the phenomenological model and the one derived from the ESR moments is minimal and does not exceed 1.5\% at any polar angle $\theta$ (see Fig.~\ref{Fig5}).

\end{document}